# Technical report: CSVM dictionaries.

## Extension of CSVM-1 specification: definition, usages, Python toolkit.

Frédéric Rodriguez [a,b,*]

[a] CNRS, Laboratoire de Synthèse et Physico-Chimie de Molécules d'Intérêt Biologique, LSPCMIB, UMR-5068, 118 Route de Narbonne, F-31062 Toulouse Cedex 9, France.
[b] Université de Toulouse, UPS, Laboratoire de Synthèse et Physico-Chimie de Molécules d'Intérêt Biologique, LSPCMIB, 118 route de Narbonne, F-31062 Toulouse Cedex 9, France.

## Abstract

CSVM (CSV with Metadata) is a simple file format for tabular data. The possible application domain is the same as typical spreadsheets files, but CSVM is well suited for long term storage and the inter-conversion of RAW data. CSVM embeds different levels for data, metadata and annotations in human readable format and flat ASCII files. As a proof of concept, Perl and Python toolkits were designed in order to handle CSVM data and objects in workflows. These parsers can process CSVM files independently of data types, so it is possible to use same data format and parser for a lot of scientific purposes.
CSVM-1 is the first version of CSVM specification, an extension of CSVM-1 for implementing a translation system between CSVM files is presented in this paper. The necessary data used to make the translation are also coded in another CSVM file. This particular kind of CSVM is called a CSVM dictionary, it is also readable by the current CSVM parser and it is fully supported by the Python toolkit. This report presents a proposal for CSVM dictionaries, a working example in chemistry, and some elements of Python toolkit usable to handle these files.

## Keywords

CSVM; Tabular Data; Python; Specification; Data Conversion; Open Format; Open Data; Open Format; Dictionaries; Canonical Data Model.

## Status

As CSVM itself, the dictionaries extension shown in this document must be considered as an Open format.

* *Corresponding authors.*
CNRS, Laboratoire de Synthèse et Physico-Chimie de Molécules d'Intérêt Biologique, LSPCMIB, UMR-5068, 118 Route de Narbonne, F-31062 Toulouse Cedex 9, France.
Tel.: þ33 (0) 5 61556486; fax: þ33 (0) 5 61556011.
E-mail address: *Frederic.Rodriguez@univ-tlse3.fr* (F. Rodriguez).

# 1. Definition of a CSVM dictionary

The CSVM dictionary 1) must use the same basis than a CVM file defined by CSVM-1 specification; 2) must be processed by the same parser that a common CSVM file and 3) must embed all the information needed to transform a CSVM file in another.
We show in this section how we can define a translation set, encode it in a CSVM dictionary and how the translation set is used to transform or normalize data of data CSVM file.

## 1.1. Data structure of a CSVM dictionary

The following CSVM file shown below is a chemical inventory table limited to 6 rows and 5 columns for simplicity. This table codes for: a rank number (`numero`), a chemical structure (`fichier`), a molecular weight (`masse_exacte`), a common molecule name (`nom`), an amount (i.e. *g* or *mg* of product) of chemical in laboratory (`vrac`) :

*Figure 1. - CSVM file for a chemical inventory.*

```
1	af01.mol	181.19293	Tyrosine	10	...
5	af02.mol	155.15753	Histidine	20	...
2	af03.mol	204.23049	Tryptophane	20	...
3	af04.mol	115.13298	Proline	12	...
4	af05.mol	267.24621	Adenosine	0	...
6	af06.mol	661.90791	Ph-Choline	300	...

#TITLE	Chemical inventory
#HEADER	numero	fichier_mol	masse_exacte	nom	vrac	...
#TYPE	NUMERIC	TEXT	NUMERIC	TEXT	NUMERIC	...
#WIDTH	10	50	50	100	10	...
#META	Get only the 5 first columns for simplicity
```

The #HEADER, #TYPE, #WIDTH define a first system (SYS1) with particular types and column names, and this file is used as a data file.
Now we want to transfer the data in another system with different naming conventions (ie. to prepare import to a RDBMS), we call it SYS2 (second system). To do this task we have defined a new CSVM file called the dictionary:

*Figure 2. - CSVM dictionary for the table shown in Figure 1.*

```
numero	ID	number	#NUMERIC	#10
nom	identificateur	name	#TEXT 	#100
fichier_mol	MOLSTRUCTURE	molfile	#TEXT	#50
masse_exacte		mol.weight	#NUMERIC	#50
vrac	vrac	qtity	#NUMERIC	#10
...

#TITLE	CSVM dictionary for SYS1, SYS2, SYS_UK
#HEADER	SYS1	SYS2	SYS1_UK	#TYPE	#WIDTH
#TYPE	TEXT	TEXT	TEXT	#TEXT	#TEXT
#WIDTH	50	50	50	#50	#50
```

We see that obviously this is also a CSVM file. The only difference is that some keywords with the # character are present in the two last columns of data and metadata blocks. These two columns (order: 4,5 from left to right) are used to store data types of translation sets, while data columns (order: 1,2,3) are used to store the translation set data itself.

This CSVM dictionary stores the columns names used in SYS1 (orange column below) and expected in SYS2 (pink) or another name space/set (green). The dictionary stores also the #TYPE and #WITH fields of each name set (blue columns) :

*Table 1. – CSVM dictionary (Figure 2.) shown as colored table.*

| | | | | | |
|---|---|---|---|---|---|
| numero | ID | number | #NUMERIC | #10 | |
| nom | identificateur | name | #TEXT | #100 | |
| fichier_mol | MOLSTRUCTURE | molfile | #TEXT | #50 | |
| masse_exacte | - | mol.weight | #NUMERIC | #50 | |
| vrac | crac | qtity | #NUMERIC | #10 | |
| | | | | | |
| #TITLE | CSVM dictionary for SYS1, SYS2 and SYS1_UK | | | | |
| #HEADER | SYS1 | SYS2 | SYS1_UK | #TYPE | #WIDTH |
| #TYPE | TEXT | TEXT | TEXT | #TEXT | #TEXT |
| #WIDTH | 50 | 50 | 50 | #50# | #50 |

So we have here all data needed to convert a CSVM file using SYS1 in another naming space because we have the #HEADER, #TYPE, #WIDTH values of SYS1 and SYS2 in this dictionary. The case of a column (#HEADER value) named TYPE or WIDTH is possible, so have chosen to mark type and width metadata in data block (blue columns) with # character to avoid confusions.

*A CSVM dictionary could be defined as a way to store Metadata of n CSVM files in the data block of another CSVM file.*

## 1.2. Translation sets

If we read the first row dictionary file, we find a value '`numero`' that is the contents of #HEADER first column in data file. The second and third values of the row ('ID' and '`number`') are alternate values of '`numero`'. If we read the first column of dictionary file, we find '`numero`', '`nom`', '`fichier_mol`', '`masse_exacte`', '`vrac`'. All are values used in data file and corresponding to SYS1, we call this ensemble a translation set. The previous dictionary defines 3 translation sets :

- The SYS1 set : [numero, nom, fichier_mol, masse_exacte, vrac]
- The SYS2 set : [ID, indentificateur, MOLSTRUCTURE, , vrac]
- The SYS1_UK set : English translation of SYS1 [`number`, `name`, `molfile`, `mol.weight`, `qttity`]

## 1.3. Guidelines for Standards and data transformation

So we can use one translation set defined in a dictionary to change values of #HEADERS of a CSVM file/object. But what about the data of target CSVM file ?
First, take in account that the perimeter of CSVM-1 covers only the syntax of a CSVM file, not the value of keywords. The recommendation about values of #TYPE fields (i.e. TEXT, NUMERIC …) must be considered only as a good practice. So, it is possible to include a lot of values for #TYPE and #WIDTH keywords, typically information to make data conversions.
The previous CSVM table shows that only one column #TYPE or #WIDTH is included in the dictionary for the all 3 translation sets. This is what we call a Standard, with this dictionary it is possible to convert all data type of each CSVM processed into the units defined by the standard.
Consider a current case in a laboratory, some columns of a CSVM file prepared by a first scientific team (encoded by translation set TEAM1) must have column names changed before data transfer to another team (translation set TEAM2). But the second team uses also different units (i.e. mass concentration unit rather than molar units).

The transformation of data units into a standard is resumed by the following figure:

*Figure 3. - Schematics of transformation of a CSVM file using a CSVM dictionary.*

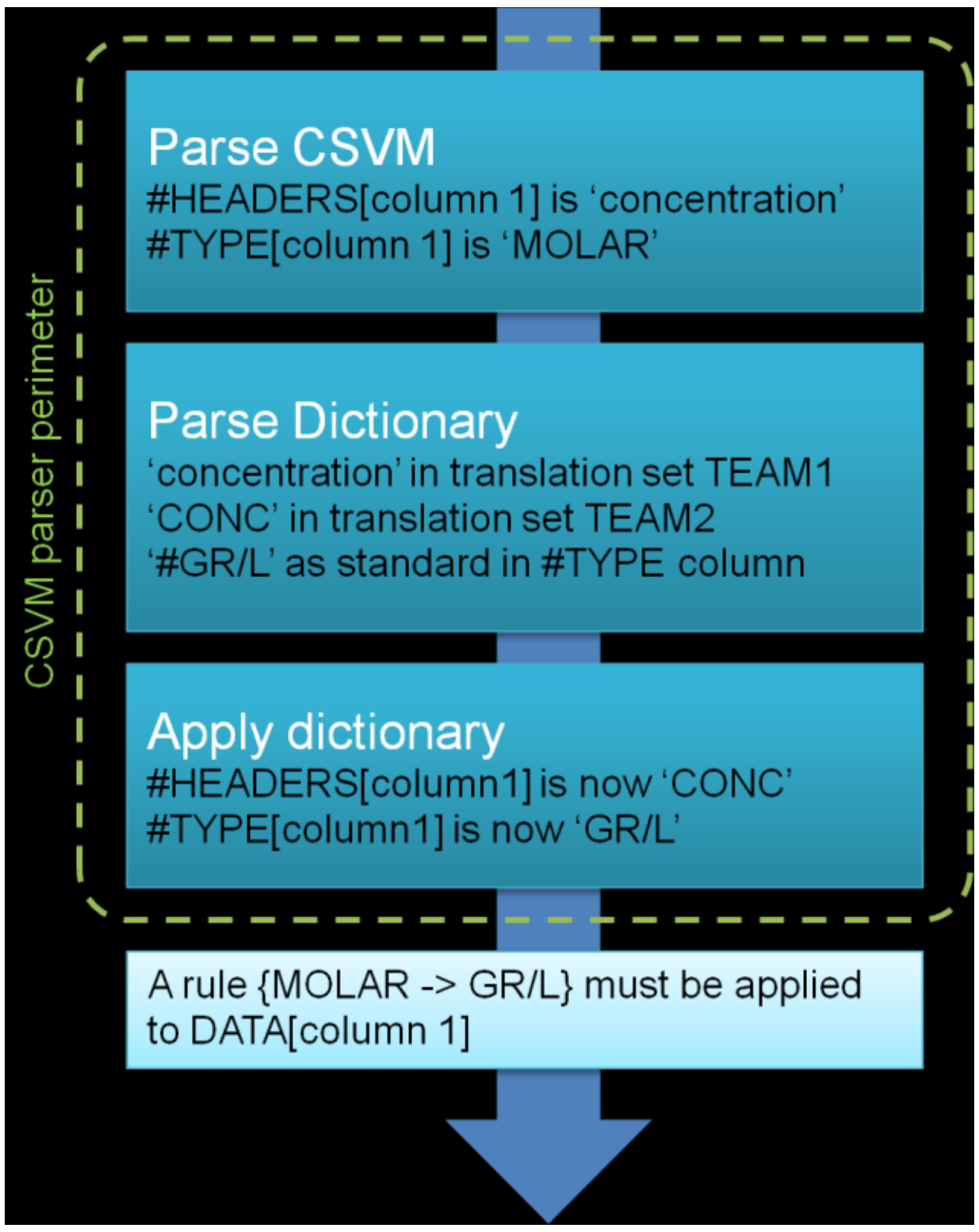

A software component knowing a rule to transform $Mol.l^{-1}$ in g/L (gram per liter) could make the data transformation, but this operation must be let out of CSVM parser's range. The corresponding coding part of dictionary for this #HEADER is given by the Table 2.

*Table 2. – Data normalization using a standard as data type in a CSVM dictionary.*

| | | | | |
|---|---|---|---|---|
| concentration | CONC | #GR/L | #10 | |
| #TITLE | Example | | | |
| #HEADER | TEAM1 | TEAM2 | #TYPE | #WIDTH |
| #TYPE | TEXT | TEXT | #TEXT | #TEXT |
| #WIDTH | 50 | 50 | #50 | #50 |

## 1.4 Guidelines for data transformation

Now if a data conversion (not the Standard) is planned for a particular column of a CSVM table ? In CSVM's context, different approaches are available. In example, if a CSVM dictionary is used [1], it is possible to add columns (TEAM1_UNITS, TEAM2_UNITS) coding for local units used by TEAM1 and TEAM2 as shown in the following table :

[1] If a dictionary is not used, the programmer could make all transformations on the target CSVM using information picked manually in another CSVM and organized in columns or rows (this approach is often used when the transformations/rules defined in dictionary must be applied sequentially). For more complex case, it could be interesting to split information in more than one CSVM file (dictionaries of common files).

*Table 3. – Adding specific columns for multi-standards.*

| | | | | | | |
|---|---|---|---|---|---|---|
| concentration | MOLAR | CONC | GR/L | #NUMERIC | #10 | |
| | | | | | | |
| #TITLE | Example | | | | | |
| #HEADER | TEAM1 | TEAM1_UNITS | TEAM2 | TEAM2_UNITS | #TYPE | #WIDTH |
| #TYPE | TEXT | TEXT | TEXT | TEXT | #TEXT | #TEXT |
| #WIDTH | 50 | 10 | 50 | 10 | #50 | #50 |

In the example shown in Table 3 no Standard is defined, but it is also possible to add one. It is also possible to extend this approach to #WIDTH columns. In dictionary file, the columns defined by values #TEAM1_UNITS and #TEAM2_UNITS are not used by the conversion functions defined in toolkits. The reason is simple: the #TEAM1_UNITS and #TEAM2_UNITS are not defined as #HEADER keywords in data files. But it is not a problem for making conversions outside the CSVM parser range, the dictionary could be read as a CSVM object and its contents used as it is a common CSVM file.

*A CSVM dictionary could also used to save knowledge about a data space of CSVM files, including supplemental fields.*

## 1.5 Lost words in translation sets

The first table (Table 1.) shows that keyword defined by '`masse_exacte`' in first translation set (SYS1 set) is not defined in SYS2 set (the char '`-`' is used here to mark an empty cell).
No particular processing is done about that, because the CSVM parser doesn't make interpretation about data or metadata values. Holes in tables are very often found in real data and can be taken in account, if a #HEADER value is not defined in the translation set, the corresponding column can be suppressed (or not) in the resulting CSVM file [2] after CSVM dictionary application.
For advanced uses, it is also possible to make a union between two CSVM files in order to restore all missing empty columns (a specific CSVM file can be forged and used as a data mask). Unions, Intersections, Additions of tables are implemented in CSVM Python toolkit, given an example:

```
print "\n*** Test CSVM files with equivalent, different, missing columns"
c1 = csvm_ptr()
c1 = csvm_ptr_read_extended_csvm(c1,"test/test1.csvm","\t")
c2 = csvm_ptr()
c2 = csvm_ptr_read_extended_csvm(c2,"test/test2.csvm","\t")
print "=> Compute INTERSECTION"
r = csvm_ptr_intersect(c1, c2)
if (r != None):
    r.csvm_ptr_dump(0,0)
    r.csvm_ptr_clear()
else:
    print "None data found"
print "\n=> compute UNION"
r = csvm_ptr_union(c1, c2)
r.csvm_ptr_dump(0,0)
r.csvm_ptr_clear()
c1.csvm_ptr_clear()
c2.csvm_ptr_clear()
```

Using CSVM paradigm, table unions or tables intersection are easy to implement, even if the two files have not exactly the same number of columns or columns names (an intermediate step using CSVM dictionaries could be added to the process).

---

[2] A typical use is the generation of a subset of RAW data (one or n CSVM files) prior to a RDBMS import.

## 1.6 Annotation of dictionaries

Dictionaries are CSVM files and can be annotated in the same way that standard CSVM. Remark lines tagged by a # character in data block (or metadata block) are used. The only rule is that a CSVM keyword (#TITLE, #HEADER, #TYPE, #WIDTH, #META) cannot be used (but combinations such as # TITLE, #_HEADER, # HEADER, ##HEADER … are allowed). The level of annotation of a CSVM file can be very high because the corresponding rows are not taken in account by the CSVM parser.
The following figure shows annotation of a CSVM file, in the case of a commented parameter's set for doing calculations using AMBER[3] [4] package.

*Figure 4. – Example of CSVM annotations inside the data block. The image is truncated after the 80$^{th}$ column (vertical grey line at right). Some lines (7, 19, 21, 31) are wrapped by the text editor [5].*

```
 1
 2  # IMIN Flag to run minimization.
 3  #===============================================================================
 4  #= 0 No minimization (only do molecular dynamics; default).
 5  #= 1 Perform minimization (and no molecular dynamics).
 6  #= 5 Read in a trajectory for analysis.
 7  # Although sander will write energy information in the output files (using ntpr),
    to post-process a set of structures using a different energy function than was usec
    replaced with a continuum model. When imin is set to 5 sander will expect to read a
    described in the mdin file for each of the structures in the trajectory file. The f
    maxcyc=1000, sander will minimize each structure in the trajectory for 1000 steps a
    to extract the energies of each of the coordinate sets in the inptraj file
 8  1	&cntrl	imin
 9
10  # MAXCYC The maximum number of cycles of minimization. Default 1
11  #===============================================================================
12  1	&cntrl	maxcyc
13
14  # NCYC If NTMIN is 1 then the method of minimization will be switched from steepest
15  1	&cntrl	ncyc
16
17  # NTB Periodic boundary.
18  #===============================================================================
19  # If NTB .EQ. 0 then a boundary is NOT applied regardless of any boundary conditior
    will be used. Options for constant pressure are described in a separate section be
20  # If NTB .NE. 0, there must be a periodic boundary in the topology file. Constant p
21  # For a periodic system, constant pressure is the only way to equilibrate density
    when solvent molecules are subtracted which can aggregate into "vacuum bubbles" in
    every system needs to be equilibrated at constant pressure (ntb=2, ntp>0) to get tc
    temperature, before turning on constant pressure.
22  1	&cntrl	ntb
23
24  # IGB Flag for using the generalized Born or Poisson-Boltzmann implicit solvent moc
25  #===============================================================================
26  # See Sections 6.1 and 6.2 for information about using this option. Default is 0
27  1	&cntrl	igb
28
29  # CUT This is used to specify the nonbonded cutoff, in Angstroms.
30  #===============================================================================
31  # For PME, the cutoff is used to limit direct space sum, and the default value of 8
    here a larger value than the default is generally required. A separate parameter (F
    summation involved in calculating the effective Born radii, see the generalized Bor
32  1	&cntrl	cut
33
34  #TITLE	CNA minimization mask
35  #HEADER	BNUM	BTYPE	KEY
36  #TYPE	NUMERIC	TEXT	TEXT
37  #WIDTH	10	10	10
38  #META	03/June/07
```

[3] The Amber Molecular Dynamics Package - http://ambermd.org/

[4] Watson-Crick base-pairing properties of nucleic acid analogues with stereocontrolled alpha and beta torsion angles (alpha,beta-D-CNAs). C. Dupouy, N. Iché-Tarrat, M.P. Durrieu, F. Rodriguez, J.M. Escudier, A. Vigroux. Angewandte Chemie (2006) 45:22, 3623-3627.

[5] Scintilla and SciTE - http://www.scintilla.org/

## 2. A working example in Chemistry

A SDF [6] file is a collection of $n$ data blocks. For each block, one can find a molecules (coded using MDL Molfile format) and [0..$m$] descriptors stored as key/values pairs, in another terms, it is a serialized molecular data table. Given a such molecular collection [7] [8], we want to transform the descriptor space: removing or renaming some of them in order to export the collection to a chemical RDBMS.

*Figure 5. – Molecular collection (displayed using a SDF viewer [9] ). The molecular formulas for each compound are shown at top part of each record. The descriptors (key/values pairs) are shown in bottom.*

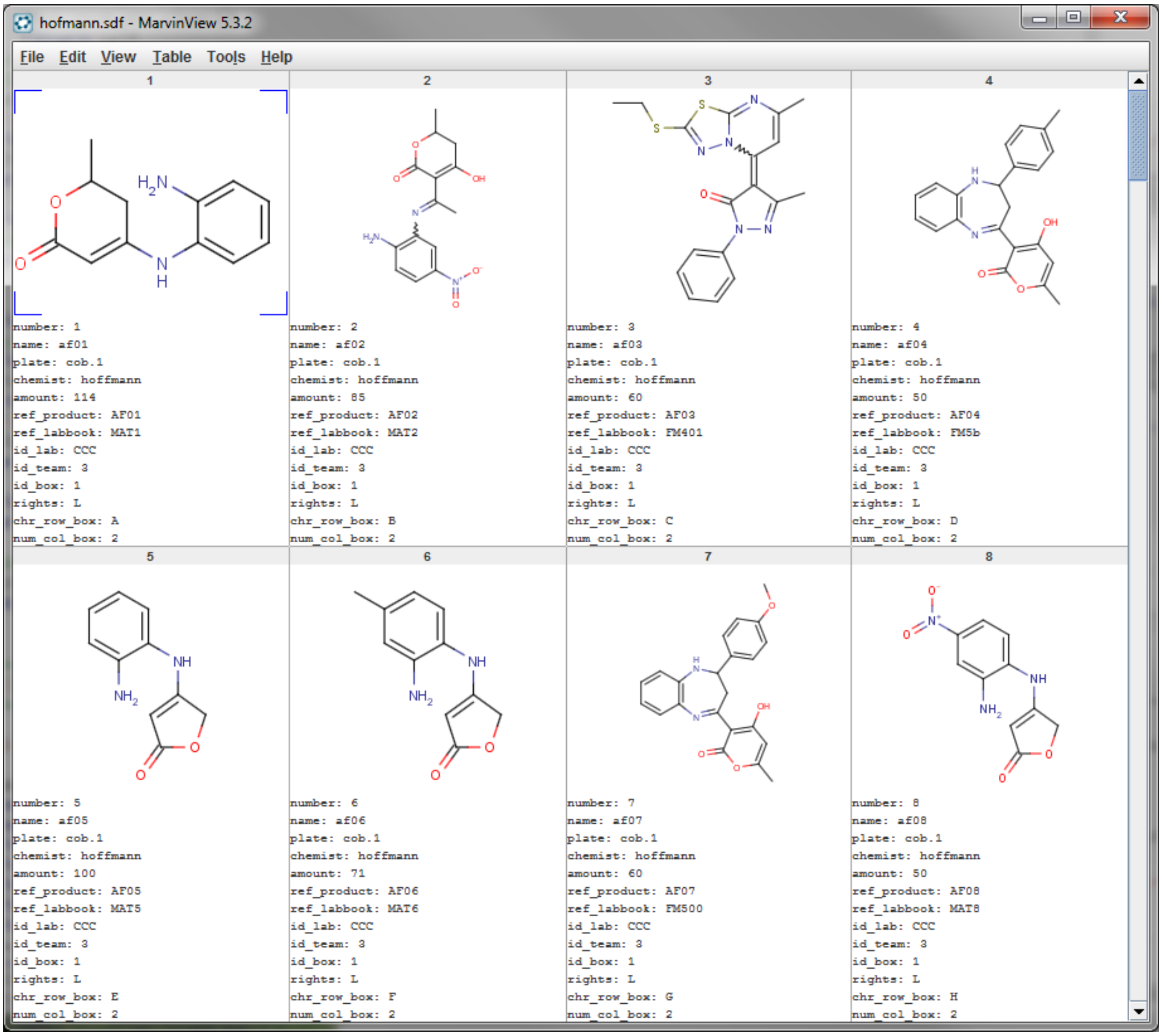

The corresponding CSVM table has one row for each compound and a column for each descriptor key (`number`, `name`, `plate`, `chemist`, `amount`, `ref_product`, `ref_labbook`, `id_lab`, `id_team`, `id_box`, `rights`, `chr_row_box`, `num_col_box`). Another column is used to store the 2D chemical structure at SMILES [10] format.

---

6 Chemical table file [Wikipedia] - http://en.wikipedia.org/wiki/Chemical_table_file

7 L. Hammal, S. Bouzroura, C. André, B. Nedjar-Kolli, P. Hoffmann (2007) Versatile Cyclization of Aminoanilino Lactones: Access to Benzotriazoles and Condensed Benzodiazepin-2-thione. *Synthetic Commun.*, 37:3, 501-511.

8 M .Fodili, M. Amari, B. Nedjar-Kolli, B. Garrigues, C. Lherbet, P. Hoffmann (2009) Synthesis of Imidazoles from Ketimines Using Tosylmethyl Isocyanide (TosMIC) Catalyzed by Bismuth Triflate. *Lett. Org. Chem.*, 6, 354-358.

9 ChemAxon marvinView - http://www.chemaxon.com/products/marvin/marvinview/

10 Chemical file format [Wikipedia] - http://en.wikipedia.org/wiki/Chemical_file_format

The following image shows the end of data block of the CSVM file and metadata block:

*Figure 6. – The molecular collection of Figure 4. encoded in a CSVM table, details only: last compounds, some rows [73 .. 80] are cut off at right. Tabs are used as field separator and are shown as red arrows.*

```
66	af66	cob.1	hoffmann	56	AF66	MAS35	CCC	03	01	L	B	10	10 11 5 9 7 3 4 6 8 2 1	C1(C(=O)SC(=N1)N)CC(=O)N
67	af67	cob.1	hoffmann	50	AF67	MAS36	CCC	03	01	L	C	10	11 12 6 10 8 3 4 7 9 2 5 1	C1(C(=O)SC(=N1)N)CC(=O)NC
68	af68	cob.1	hoffmann	50	AF68	MAS37	CCC	03	01	L	D	10	9 12 8 13 10 7 2 3 11 6 5 4 1	C1(C(=O)SC(=N1)N)CC(=O)NCC
69	af69	cob.1	hoffmann	80	AF69	MAS38	CCC	03	01	L	E	10	13 9 5 4 5 9 15 12 8 14 7 2 3 6 11 1 10	c1(ccccc1)NC(=O)CC1C(=O)SC(=N1)N
70	af70	cob.1	hoffmann	50	AF70	A23	CCC	03	01	L	F	10	8 11 7 12 9 5 4 3 10 6 2 1	C1(C(=O)SC(=N1)NC)CC(=O)N
71	af71	cob.1	hoffmann	58	AF71	A21	CCC	03	01	L	G	10	12 13 7 11 9 4 6 8 10 3 2 5 1	C1(C(=O)SC(=N1)NCC)CC(=O)N
72	af72	cob.1	hoffmann	50	AF72	A22	CCC	03	01	L	H	10	13 9 5 4 5 9 15 12 8 14 7 3 10 6 11 2 1	c1(ccccc1)NC1=NC(C(=O)S1)CC(=O)N
73	af73	cob.1	hoffmann	33	AF73	FM421	CCC	03	01	L	A	11	20 16 10 15 11 17 19 18 12 8 6 5 13 3 9 14 2 7 4 1	n12c(cc(nc1cc1c2cccc1)C)CC(=O)OCC
74	af74	cob.1	hoffmann	45	AF74	FM245	CCC	03	01	L	B	11	21 9 6 10 18 13 11 1 19 12 20 17 14 8 16 7 5 15 7 4 5 3 2	n12ncsc1nc(C)c/c/2=C/1\C(
75	af75	cob.1	hoffmann	50	AF75	FM255	CCC	03	01	L	C	11	22 9 14 11 19 13 10 1 20 12 21 18 15 8 17 7 6 16 7 5 6 4 3 2	n12nc(sc1nc(C)c/c/2=C
76	af76	cob.1	hoffmann	50	AF76	FM414	CCC	03	01	L	D	11	22 9 14 11 19 13 10 1 20 12 21 18 15 8 17 7 6 16 7 5 6 4 3 2	n12nc(sc1nc(C)c/c/2=C
77	af77	cob.1	hoffmann	45	AF77	FM410	CCC	03	01	L	E	11	18 8 6 5 6 8 24 20 17 9 19 26 21 11 15 13 22 23 25 12 7 16 10 14 4 2 3 1	c1(ccccc1
78	af78	cob.1	hoffmann	35	AF78	FM406	CCC	03	01	L	F	11	19 9 6 5 6 9 25 21 18 10 20 27 22 12 16 14 23 24 26 13 8 17 11 15 4 2 3 7 1	c1(ccccc1
79	af79	cob.1	hoffmann	43	AF79	FM267	CCC	03	01	L	G	11	21 16 7 6 7 16 26 22 27 11 20 9 17 10 19 23 25 14 18 15 24 12 8 13 2 5 1 4 3	c1(cc
80	af80	cob.1	hoffmann	45	AF80	FM257	CCC	03	01	L	H	11	8 6 5 7 15 16 3 12 14 1 19 11 2 21 13 17 18 4 9 20 10	s1cccc1Cn1c(c(nc1)C)c1c(O)cc(

#TITLE	
#HEADER	number	name	plate	chemist	amount	ref_product	ref_labbook	id_lab	id_team	id_box	rights	chr_row_box	num_col_box	OpenBabel Symmetry Classes	smi
#TYPE	TEXT	TEXT	TEXT	TEXT	TEXT	TEXT	TEXT	TEXT	TEXT	TEXT	TEXT	TEXT	TEXT	TEXT	TEXT
#WIDTH	10	10	10	10	10	10	10	10	10	10	10	10	10	10	10
#META	
```

A first CSVM dictionary (*dict1,* Figure 7) will be used to filter CSVM columns of the molecular collection (Figure 6). In *dict1*, the columns 1-2-3 of data block, list the keywords of CSVM files allowed in corresponding translation sets LOCAL, LOCAL2, CN. If the translation occurs from LOCAL to LOCAL2, all columns of the molecular table that are named `num_col_box` will be renamed to `ccol`.

*Figure 7. – A CSVM dictionary (dict1).*

```
number	ID	ID	#NUMERIC	#10
name	identificateur	identificateur	#TEXT	#50
file_mol	MOLSTRUCTURE	__DEL__	#FILE	#50
amount	vrac	vrac	#NUMERIC	#10
plate	plaque	plaque	#TEXT	#10
chemist	laboratoire	__DEL__	#TEXT	#50
remarks	observations	__DEL__	#TEXT	#100
ref_product	reference_produit	__DEL__	#TEXT	#20
ref_labbook	reference_cahier	__DEL__	#TEXT	#20
id_lab	clab	__DEL__	#TEXT	#10
id_team	ceq	__DEL__	#NUMERIC	#10
id_box	cbox	__DEL__	#NUMERIC	#10
rights	cleg	__DEL__	#TEXT	#10
chr_row_box	clig	__DEL__	#TEXT	#10
num_col_box	ccol	__DEL__	#NUMERIC	#10
smi	smi	smi	#TEXT	#10
mdl	mdl	mdl	#TEXT	#50
date	date	__DEL__	#TEXT	#10
OpenBabel Symmetry Classes	__DEL__	__DEL__	#TEXT	#50

#TITLE	Headers dicts to use with transforms 
#HEADER	LOCAL	LOCAL2	CN	#TYPE	#WIDTH
#TYPE	TEXT	TEXT	TEXT	#TEXT	#TEXT
#WIDTH	50	50	50	#50	#50
```

Some values in *dict1*.CN are set to `__DEL__` string: if translation occurs from LOCAL/LOCAL2 to CN, all columns tagged with `__DEL__` will be deleted in the resulting collection:

*Figure 8. – Molecular collection after application of dict1.CN filter.*

```
66	af66	cob.1	56	C1(C(=O)SC(=N1)N)CC(=O)N
67	af67	cob.1	50	C1(C(=O)SC(=N1)N)CC(=O)NC
68	af68	cob.1	50	C1(C(=O)SC(=N1)N)CC(=O)NCC
69	af69	cob.1	80	c1(ccccc1)NC(=O)CC1C(=O)SC(=N1)N
70	af70	cob.1	50	C1(C(=O)SC(=N1)NC)CC(=O)N
71	af71	cob.1	58	C1(C(=O)SC(=N1)NCC)CC(=O)N
72	af72	cob.1	50	c1(ccccc1)NC1=NC(C(=O)S1)CC(=O)N
73	af73	cob.1	33	n12c(cc(nc1cc1c2cccc1)C)CC(=O)OCC
74	af74	cob.1	45	n12ncsc1nc(C)c/c/2=C/1\C(=O)N(N=C1C)c1ccccc1
75	af75	cob.1	50	n12nc(sc1nc(C)c/c/2=C/1\C(=O)N(N=C1C)c1ccccc1)C
76	af76	cob.1	50	n12nc(sc1nc(C)c/c/2=C/1\C(=O)N(N=C1C)c1ccccc1)S
77	af77	cob.1	45	c1(ccccc1)N1C(=O)/C(=c\2/n3c(nc(c2)C)sc2c3ccc(c2)C)/C(=N1)C
78	af78	cob.1	35	c1(ccccc1)N1C(=O)/C(=c\2/n3c(nc(c2)C)sc2c3ccc(c2)OC)/C(=N1)C
79	af79	cob.1	43	c1(ccccc1)n1c(c(c(n1)C)C(=O)/C=C(\Nc1sc2c(n1)c(ccc2)C)/C)O
80	af80	cob.1	45	s1cccc1Cn1c(c(nc1)C)c1c(O)cc(C)oc1=O

#TITLE	
#HEADER	ID	identificateur	plaque	vrac	smi
#TYPE	TEXT	TEXT	TEXT	TEXT	TEXT
#WIDTH	10	10	10	10	10
#META	
```

The corresponding SDF file is shown in Figure 9. The molecular formulas are regenerated and some keywords are missing or are renamed (`number -> ID`, `name -> Identificateur`, `plate -> plaque`, `amount -> vrac`) in the new SDF file.

*Figure 9. – Molecular collection after filtering by dict1.CN set.*

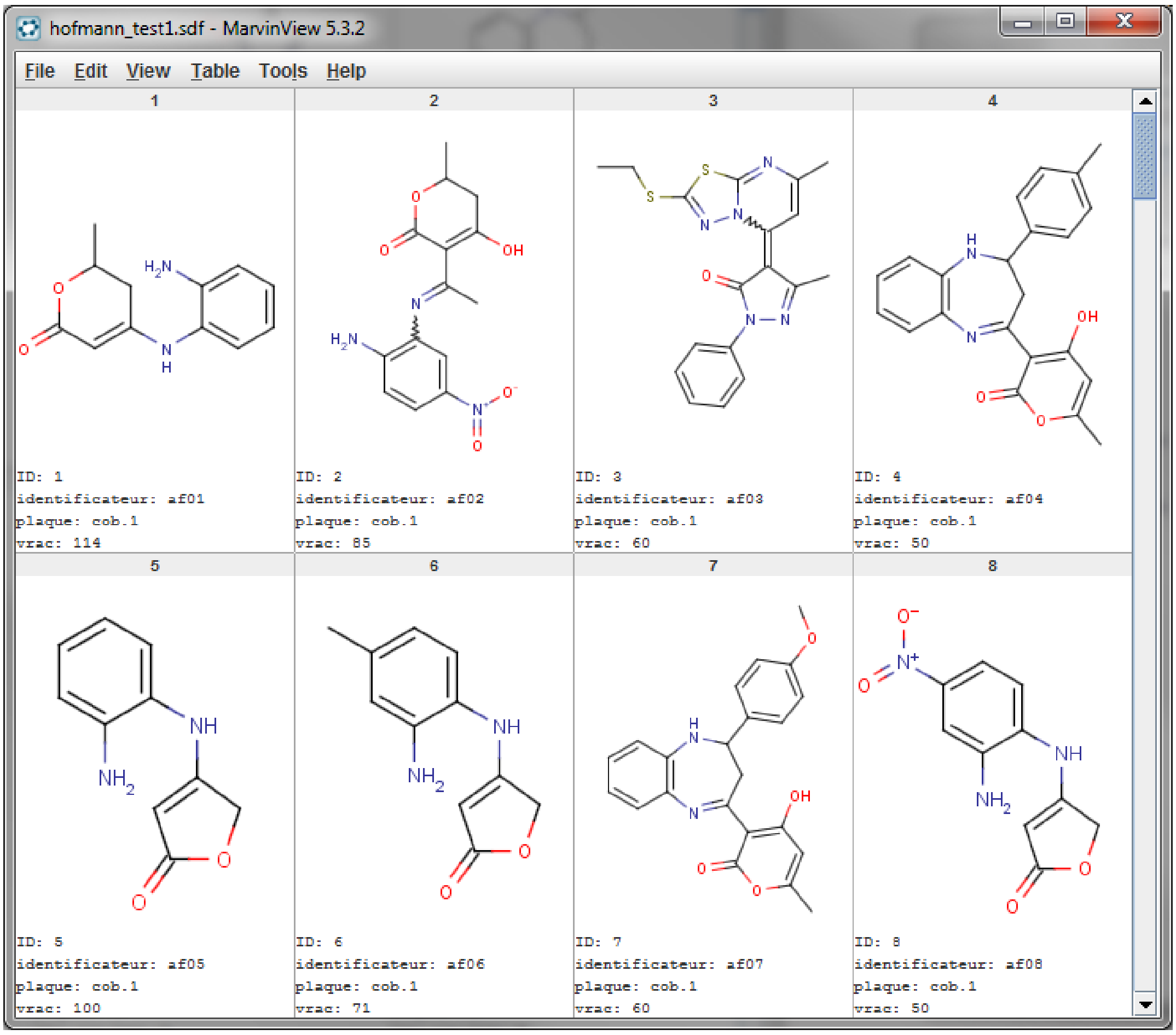

This example shows how to use CSVM as intermediate format in order to make operations (here a simple filtering) on tables. This kind of operation is very often done in a lot of scientific fields and using CSVM let us integrate table in scientific workflows and normalize data.

CSVM's paradigm permits a real data management and storage of RAW data. Eventually these RAW data will be integrated in a RDBMS, if this job is done by another team, the CSVM files are enough documented, and the producer's team will be very little demand. If the RAW data will not be integrated in a RDBMS: all that is needed (metadata, annotations) to use these files some years later is embedded inside (event is a CSVM parser is not available, all information is human readable).

# 3. Python toolkit

The CSVM dictionaries are fully supported in the Python toolkit for CSVM, the following code was used for the translation shown in previous section:

*Code 1. – Filtering process.*

```
if __name__ == '__main__':
   from build.file import file_file2str, file_str2file, file_cleanpath
   print "*** CSVM dictionary test"
   print
   print "*** TEST1: using a dictionary in which __DEL__ are included."
   print "*** strong mode IS NOT used"
   print "\n=> A new blank CSVM object"
   c = csvm_ptr()
   print "\n=> Populates it with a CSVM file ... "
   c = csvm_ptr_read_extended_csvm(c, file_cleanpath("test/hoffmann.csvm"), "\t")
   c.csvm_ptr_dump(0,0)
   print "\n=> Apply a filter using a dictionary"
   dict_file = file_cleanpath("test/dictionary_test1.csvm")
   print "\n=> dictionary file is [%s]\n" % (dict_file)
   c = csvm_dict_file_filter(c, dict_file, 'CN', 0)
   print "\n=> resulting output"
   c.csvm_ptr_dump(0,0)
   print "\n=> save new CSVM file"
   s = csvm_ptr_make_csvm(c,"\n","\t")
   file_str2file(file_cleanpath("test/hofmann_test1.csvm"), s)
   print "\n=> Clear CSVM object"
   c.csvm_ptr_clear()
```

The `csvm_ptr_read_extended_csvm` functions make a CSVM object (`c`) in memory, using a CSVM file. The method `csvm_dict_file_filter` make the filter operation. It uses the name of the *dict1* dictionary file (“`dictionary_test1.csvm`” stored in argument *dict_file*) and the translation set ‘CN’. Then the CSVM object is converted in a file and the memory cleared.
The `csvm_dict_file_filter` launch a simple function `csvm_dict_ptr_filter` to do the filtering :

*Code 2. – Basic filtering.*

```
def csvm_dict_ptr_filter(self, dict, set, delcol='__DEL__'):
   """
   The subroutine filters a CSVM file (a csvm_ptr object self) using a
   dictionary (another csvm_ptr object, dict). The HEADER values (columns names
   of CSVM structure) are translated using the set of dictionnary given as
   argument. If the HEADER values are explicitely translated to '__DEL__' value,
   the corresponding columns of CSVM structure are deleted.
   The argument set is a string used as the identifier (column name, element
   of #HEADER list) of a translate set included in dict (a csvm_ptr object).
   *** we call 'standard' this filtering mode.
   """
   if (dict.HEADER_N <= 0): return self
   dict.csvm_ptr_dump(0,0)
   if (len(set) <= 0): return self
   if ((set in dict.HEADER) == False): return self
   self = csvm_ptr_colfilter(self, dict, set)
   self = csvm_ptr_delcol(self, delcol)
   return self
```

One function does the filtering (`csvm_ptr_colfilter`) and the other (`csvm_ptr_delcol`) does the column removal.

## 3.1 Special cases: strong mode

Sometimes user cannot define if columns must be removed explicitly in a translation set, or if columns have a corresponding value in another translation set.
The Figure 10 illustrates this case for a CSVM dictionary (*dict2*) derived from previous *dict1*. The values `__DEL__` have disappeared from the table and are replaced by characters ’-‘ currently used to mark empty cells in the CSVM files.

*Figure 10. – A CSVM dictionary (dict2) without explicit column deletion.*

```
number	ID	ID	#NUMERIC	#10
name	identificateur	identificateur	#TEXT	#50
file_mol	MOLSTRUCTURE	-	#FILE	#50
amount	vrac	vrac	#NUMERIC	#10
plate	plaque	plaque	#TEXT	#10
chemist	laboratoire	-	#TEXT	#50
remarks	observations	-	#TEXT	#100
ref_product	reference_produit	-	#TEXT	#20
ref_labbook	reference_cahier	-	#TEXT	#20
id_lab	clab	-	#TEXT	#10
id_team	ceq	-	#NUMERIC	#10
id_box	cbox	-	#NUMERIC	#10
rights	cleg	-	#TEXT	#10
chr_row_box	clig	-	#TEXT	#10
num_col_box	ccol	-	#NUMERIC	#10
smi	smi	smi	#TEXT	#10
mdl	mdl	mdl	#TEXT	#50
date	date	-	#TEXT	#10
OpenBabel Symmetry Classes	-	-	#TEXT	#50

#TITLE	Headers dicts to use with transforms
#HEADER	LOCAL	LOCAL2	CN	#TYPE	#WIDTH
#TYPE	TEXT	TEXT	TEXT	#TEXT	#TEXT
#WIDTH	50	50	50	#50	#50
```

In this case another approach could be used, first the calling code similar to *Code 1* example, but using strong mode: the last argument of `csvm_dict_file_filter` function is set to 1 rather than zero.

*Code 3. – Advanced filtering.*

```
print "*** TEST2: using a dictionary without __DEL__ or meta commands."
   print "*** strong mode IS used"
   print "\n=> A new blank CSVM object"
   c = csvm_ptr()
   print "\n=> Populates it with a CSVM file ... "
   c = csvm_ptr_read_extended_csvm(c, file_cleanpath("test/hoffmann.csvm"), "\t")
   c.csvm_ptr_dump(0,0)
   print "\n=> Apply a filter using a dictionary"
   dict_file = file_cleanpath("test/dictionary_test2.csvm")
   print "=> dictionary file is [%s]" % (dict_file)
   c = csvm_dict_file_filter(c, dict_file, 'CN', 1)
   print "\n=> resulting output"
   c.csvm_ptr_dump(0,0)
   print "\n=> save new CSVM file"
   s = csvm_ptr_make_csvm(c,"\n","\t")
   file_str2file(file_cleanpath("test/hofmann_test2.csvm"), s)
   print "\n=> Clear CSVM object"
   c.csvm_ptr_clear()
```

The `csvm_dict_ptr_filter_blank` function is called and could be taken as a basis to make more advanced filtering subroutines.

*Code 4. – Evolution of csvm_dict_ptr_filter subroutine.*

```
def csvm_dict_ptr_filter_blank(self, dict, set, blank_list=['', '-'], delcol='__DEL__'):
   """
   The subroutine filters a CSVM file like csvm_dict_ptr_filter subroutine
   but using another approach to destroy columns. A csvm_ptr object (self) is
   given in input and filtered using a set (set) of a CSVM dictionary (dict).
   The argument set is a string used as the identifier (column name, element
   of #HEADER list) of a translate set included in dict (a csvm_ptr object).
   If a column name of self, is not found in the list defined by set, the
   corresponding column must be deleted in self. The argument blank_list
   is used to store the values used in cells for no data (empty string or
   a special char as '-').
   If the argument delcol is set (and its length is > 1), the destruction of
   corresponding columns (same as csvm_dict_ptr_filter) is applied before
   destruction of columns with data of blank_list.
   The values of blank_list are case sensitive and comparated in equal (not
   include) mode for security reasons about data.
   *** we call 'strong' or 'blank' this filtering mode.
   """
   if (dict.HEADER_N <= 0): return self
   if (len(set) <= 0): return self
   if ((set in dict.HEADER) == False): return self
## apply filter
   self = csvm_ptr_colfilter(self, dict, set)
   if (len(delcol) > 1):
       self = csvm_ptr_delcol(self, delcol)
```

```
## construct kwlist (only unique values in words list)
   iset = query_row_eq(dict.HEADER,set,1,0)
   if (iset < 0): return (self)
   kwlist = []
   for i in range (0, dict.DATA_R, 1):
       kwlist.append(dict.DATA[i][iset[0]])
## remove blank cells in kwlist
   for i in range (len(kwlist)-1, -1, -1):
       for j in range (0, len(blank_list), 1):
           if (kwlist[i] == blank_list[j]):
               del(kwlist[i])
## query using kwlist and column(s) removal
   dlist = query_row_not_eqsv(self.HEADER, kwlist, 'AND', 1, 0)
   if ((len(dlist) <= 0) or (dlist == None)): return self
   hlist = []
   for i in range (0, len(dlist), 1):
       hlist.append(self.HEADER[dlist[i]])
   for i in range (0, len(hlist), 1):
       self = csvm_ptr_delcol(self, hlist[i])
## done
   return self
```

The first part (## apply filter) is the same as code shown in Code 2 example. The iset variable is used to store the index of translation set selected in the *dict2*, in this case *iset = [2]* because it is the third element of #HEADER list.
The kwlist Python list (mono-dimensional array) is used to store the column names corresponding to the translation set: *kwlist = ['ID', 'identificateur', '-', 'vrac', 'plaque', '-', '-', '-', '-', '-', '-', '-', '-', '-', '-', 'smi', 'mdl', '-', '-'].*
This list is filtered using the argument blank_list of the function. With default values the empty cells or cell tagged by '–' are removed from kwlist, and now, *kwlist = ['ID', 'identificateur', 'vrac', 'plaque', 'smi', 'mdl'].*
Knowing this information, it is possible to get the column names (in this case indexes are returned) of molecular table that are not found in kwlist. And we have *dlist = [3, 5, 6, 7, 8, 9, 10, 11, 12, 13]*. The corresponding column names are stored in hlist, in this case *hlist = ['-', '-', '-', '-', '-', '-', '-', '-', '-', '-']* and all columns with names included in hlist will be deleted.

Here all columns names of molecular collection are specified in the dictionary (using a value or a '–'). In the general case, a dictionary may not have all possible column names that can be found in data files. To illustrate this remark, we can comment two rows in *dict2* dictionary, one that is included in target translation set (`plate` row) and one that is not included (`chemist`). These two rows are tagged using the '#' character in first position and will not be read by the CSVM parser (they are considered now like annotations in the data block of the dictionary):

```
#plate      plaque        plaque   #TEXT   #10
#chemist    laboratoire   -        #TEXT   #50
```

The corresponding values of intermediate variables are:

*iset = [2]*
*kwlist = ['ID', 'identificateur', '-', 'vrac', '-', '-', '-', '-', '-', '-', '-', '-', '-', 'smi', 'mdl', '-', '-']*
*kwlist = ['ID', 'identificateur', 'vrac', 'smi', 'mdl']*
*dlist = [2, 3, 5, 6, 7, 8, 9, 10, 11, 12, 13]*
*hlist = ['plate', 'chemist', '-', '-', '-', '-', '-', '-', '-', '-', '-']*

And the column '`plate`' is not added to the final molecular collection. The following lines are the dump of the corresponding CSVM object (only the ten first lines of the data block are shown):

```
=> resulting output

DUMP: CSVM info {
SOURCE test\hoffmann.csvm
CSV    CSVM
META   []
TITLE_N       1
TITLE
HEADER_N      4
TYPE_N 4
WIDTH_N       4
0      10     TEXT   {ID}
1      10     TEXT   {identificateur}
```

```
2      10     TEXT   {vrac}
3      10     TEXT   {smi}
DATA_R 80
DATA_C 4
       80     4
0      [01][af01][114][C1C(OC(=O)C=C1Nc1ccccc1N)C]
1      [02][af02][85][c1(c(ccc(c1)[N+](=O)[O-])N)/N=C(/C1=C(O)CC(OC1=O)C)\C]
2      [03][af03][60][n12nc(sc1nc(C)c/c/2=C/1\C(=O)N(N=C1C)c1ccccc1)SCC]
3      [04][af04][50][c1(ccc(cc1)C)C1Nc2c(N=C(c3c(=O)oc(cc3O)C)C1)cccc2]
4      [05][af05][100][C1(=CC(=O)OC1)Nc1ccccc1N]
5      [06][af06][71][C1(=CC(=O)OC1)Nc1ccc(cc1N)C]
6      [07][af07][60][c1(ccc(cc1)OC)C1Nc2c(N=C(c3c(=O)oc(cc3O)C)C1)cccc2]
7      [08][af08][50][C1(=CC(=O)OC1)Nc1ccc(cc1N)[N+](=O)[O-]]
8      [09][af09][60][c12c(cccc2)n(cn1)C1=CC(=O)OC(C1)C]
9      [10][af10][45][c12c(ccc(c2)C)n(cn1)C1=CC(=O)OC(C1)C]
10     [11][af11][43][c12c(cc(cc2)Cl)n(cn1)C1=CC(=O)OC(C1)C]
…
```

## 5. Conclusion and perspectives

One typical tedious task is to develop software in order to convert different data flows. The combined use of text components and CSVM dictionaries has helped us to reduce greatly the number of lines of code. The use of dictionaries rather that UNIX filters has considerably helped all users to design-debug-share filters.
All the steps of data flows can be documented: the data itself, the dictionaries, the indexes of collections (data, files), the intermediate files used in interchange flows … allowing re-using information at high level, even for RAW data.
The CSVM files are easy to generate, not only for automatic processes but also for humans. The only need is to add a metadata block at the bottom of a spreadsheet file and to save it using a ASCII/CSVM format and a well chosen field delimiter.
For these reasons CSVM seems a good candidate for being a canonical data model in a lot of applications for science and industry.

## Supporting information

Please contact corresponding author for support on the Python (Pybuild) CSVM toolkit or data format.
The CSVM-1 specification details can be found in the following reference:
G. Beyries, F. Rodriguez (2012) Technical Report: CSVM format for scientific tabular data - arXiv:1207.5711v1 [ http://fr.arxiv.org/abs/1207.5711v1 ].

## Acknowledgments

I am grateful to Dr Michel Baltas and Dr Casimir Blonski (LSPCMIB, UMR 5068 CNRS-Toulouse University) for supporting this work.
I would like to especially thank Dr Jean Louis Tichadou ("Université Paul Sabatier", Toulouse University) for helpful discussions and for the support of University Course (2006-2011) around RAW data questions in experimental or environmental sciences.
I thank all collaborators in different laboratories which shared data, and help to the development the format's usage, especially Dr Pascal Hoffmann (LSPCMIB, UMR 5068 CNRS-Toulouse University) which has provided the chemical data used as example in section 2 of this manuscript.
I would like to thank the CNRS, the "Université Paul Sabatier" for their financial support.